\documentclass[prd, twocolumn, superscriptaddress, nofootinbib]{revtex4}
\usepackage{graphicx, epsfig}

\def\la{\mathrel{\mathpalette\fun <}}

\def\fun#1#2{\lower3.6pt\vbox{\baselineskip0pt\lineskip.9pt
  \ialign{$\mathsurround=0pt#1\hfil##\hfil$\crcr#2\crcr\sim\crcr}}}

\begin{document}

\title{Importance of Supernovae at $z>1.5$ to Probe Dark Energy}
\author{Eric V. Linder}
\affiliation{Physics Division, Lawrence Berkeley National Laboratory, Berkeley, CA 94720} 
\author{Dragan Huterer}
\affiliation{Department of Physics, Case Western Reserve University,
Cleveland, OH~~44106}

\begin{abstract} 
The accelerating expansion of the universe suggests that an unknown
component with strongly negative pressure, called dark energy,
currently dominates the dynamics of the universe.  Such a component
makes up $\sim70\%$ of the energy density of the universe yet has
not been predicted by the standard model of particle physics.  The
best method for exploring the nature of this dark energy is to map the
recent expansion history, at which Type Ia supernovae have proved
adept.  We examine here the depth of survey necessary to provide a
precise and qualitatively complete description of dark energy.
Realistic analysis of parameter degeneracies, allowance for natural
time variation of the dark energy equation of state, and systematic
errors in astrophysical observations all demonstrate the importance of
a survey covering the full range $0<z\la2$ for revealing the
nature of dark energy.
\end{abstract}

\preprint{CWRU-08-02}
\maketitle

\section{Introduction} 

The discovery of the acceleration of the expansion of the universe
through the Type Ia supernova distance-redshift relation is a major
development in cosmology~\cite{SCP,HIGHZ}. Exploring the expansion
history of the universe is a key aim of cosmology, producing literally
a textbook picture of the universe. Furthermore, such a map provides
key clues to the underlying physics, independent of whether this is
dark energy, higher dimensions, or an altered theory of gravitation
\cite{Li02}.

In its interpretation as arising from a universal vacuum, or dark,
energy, such a component would comprise some 70\% of the critical
density, be unclustered on subhorizon scales, and possess a
substantially negative equation of state (EOS) $w=p/\rho\lesssim -0.6$
\cite{w6}.  While these properties are unexpected from the standard
model of particle physics, it has been suggested that they can be
motivated by a number of fundamental theories
\cite{WA,RMP}. Dark energy thus poses a crucial 
mystery to unravel for the fields of high energy physics, cosmology,
and gravitation.

\begin{figure}[!ht]
\begin{center} 
\epsfig{file=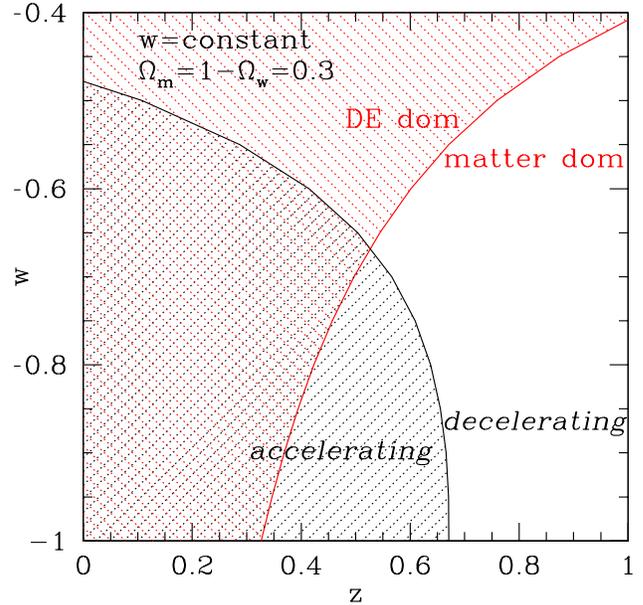, height=3.3in, width=3.4in} 
\caption{The epochs of equality between the dark energy density 
and matter and of transition from acceleration to deceleration 
are plotted vs.~dark energy equation of state.  The positively 
slanted hatching denotes the accelerating phase; the negatively 
slanted hatching shows when the dark energy density dominates over 
the matter density. Despite these 
both occurring below redshift $z\approx0.7$, dark energy can be 
probed to much higher redshift. 
} 
\label{zeq}
\end{center} 
\end{figure}

Supernovae studies, which first provided the evidence for the
acceleration, are well suited for elucidating the nature of the dark
energy~\cite{triangle,HT}. One experiment being designed specifically
to probe the accelerating universe using supernovae is the 
Supernova/Acceleration Probe (SNAP \cite{snap}). 
At an initial theoretical glance, the redshift range over
which this exploration is most easily done seems simple to understand:
the energy density dominance and dynamical influence (accelerating
power) of dark energy enters at redshifts $z\lesssim 0.7$ (see
Fig.~\ref{zeq}).  Moreover, an idealized perturbative, or Fisher
matrix, calculation shows that the ``sweet spot'' of sensitivity to
the equation of state $w$ lies at $z\approx0.3$~\cite{HT,WA,HS}. So why
are observations at $z>1$ necessary for characterizing the dark
energy?

The answer lies in the breakdown of the ideal case: 
\begin{itemize} 
\item Cosmological degeneracies 
\item Dark energy model degeneracies 
\item Systematic errors 
\end{itemize} 
The required survey depth depends on the rigor of our scientific
investigation, how much we are willing to assume about the other
parameters entering into the determination of the dark energy equation
of state.  One could estimate a false precision without knowing how
accurate, i.e.~biased, the result is.  We label this blind trust by
three heresies\footnote{The authors in no way advocate burning at the
stake.}, and here aim to demonstrate their insidious effects through
simple illustrations rather than mathematical arguments.

\section{Heresy by Word: Dark Energy is only seen at low \lowercase{$z$}}

\begin{figure}[!ht]
\begin{center} 
\epsfig{file=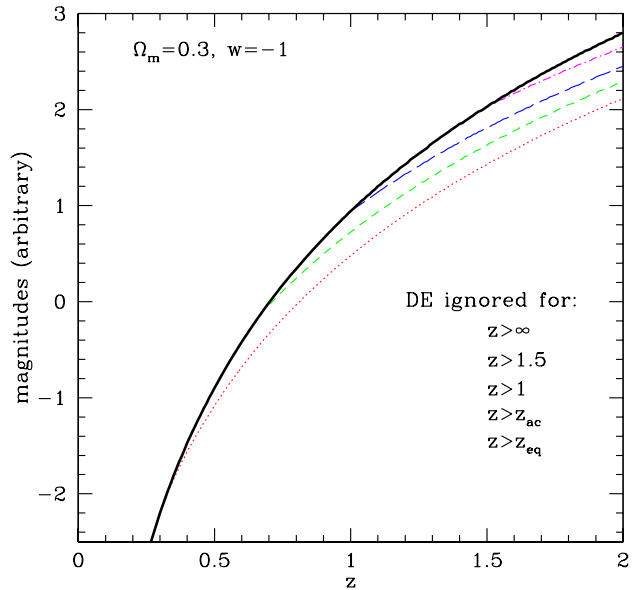, height=3.3in, width=3.4in} 
\caption{Dynamical influence of dark energy persists substantially 
beyond the redshifts of equality $z_{eq}$ or the acceleration-deceleration 
transition $z_{ac}$. The curves show how the magnitude-redshift relation is 
distorted when the dark energy is ignored (i.e.\ treated as ordinary matter) 
above  different redshifts (labeled from top down).  The thickness of 
the solid black curve that fully incorporates dark energy represents 
0.02 magnitudes -- SNAP's projected sensitivity -- so dark energy 
influence remains quite detectable even at $5z_{eq}$. 
} 
\label{zeqm}
\end{center} 
\end{figure}

Acceleration of the expansion must give way as we look further into
the past to a normal, matter dominated decelerating phase so that
structure could have formed.  Observation of the turnover in the 
distance-redshift relation due to this transition provides
both a critical check on our understanding and a discriminator from
(generically monotonic) systematic effects; this requires redshifts $z>1$.  
While Fig.~\ref{zeq} shows the acceleration/deceleration
transition occurs at lower $z$, the inertia caused by the integral
nature of the distance relation prevents the turnover in the
magnitude-redshift Hubble diagram from appearing until higher redshift
\cite{MBS,Li01}.  
The turnover occurs when the EOS of the total energy density $w_T=-1/3$. 
Distinguishing between dark energy models based on their distance-redshift 
behavior depends on the difference between their $w_T(z)$, but the 
models can cross in $w_T-z$ plane.  Therefore, 
Hubble diagram curves of models may diverge only slowly with redshift.  
These effects preserve the importance of dark energy at
higher redshifts. Fig.~\ref{zeqm} illustrates the falsity of the
na\"\i ve assumption that dark energy is only important at low
redshift: dark energy has an influence, significant on the precision
scales SNAP can achieve, out beyond $z=1.5$.  A survey extending this
deep can clearly map out the transition from the accelerating to
decelerating phase, basically seeing the onset of a present day
inflation \cite{Li02,turner_riess}.

\section{Heresy by Thought: Ignoring Time Variation \lowercase{$w'$}} 

A leading candidate for the physics behind the accelerating universe
is a dynamical scalar field acting as vacuum energy.  But high energy
field theories generically predict that the equation of state of such
a dark energy -- other than the cosmological 
constant -- should vary with time.  So consideration of only
constant $w$ models severely prejudices the parameter space of
theories. Conventionally one enlarges the classes of fundamental
physics probed by including time variation to first order:
$w(z)=w_0+w'z$~\cite{coohut}. The parameter $w'$ is directly 
related to the scale length of the field potential $V'/V\equiv d\ln
V/d\phi$.   

Allowing for $w'$ has a dramatic effect on the physical 
content of the results.  Consider the analogy of the now classic 
confidence contours in the dark energy (cosmological constant) 
density vs.~matter density, 
or $\Omega_\Lambda-\Omega_M$, plane.  Finding a  
precise value of, say, $\Omega_M=0.45$, $\Omega_\Lambda=1$ -- 
purely hypothetical but consistent with current supernova data -- 
would contradict 
CMB results on flatness.  Should we interpret this as evidence 
for a radical reworking of cosmology?  Not necessarily, for the simpler 
explanation is that we unnecessarily limited the dark energy parameter 
space by forcing $w=-1$, a cosmological constant.  Such a hypothetical 
result could be equally well fit (over a redshift range $z\lesssim 1$) 
by a consistent flat model 
with $\Omega_M=0.3$, $w=-1.15$.  Analogously, confining ourselves 
to constant $w$ can skew the results from the true model containing 
a natural $w'$ term -- with a very different underlying physics. 
That is, a restricted phase space is subject to bias because of ignoring 
other parameters\footnote{Rather than calling these families of models 
degenerate, it is 
more evocative to call them congeneric: resembling in nature or action. 
This has the connotation in chemistry of a molecule that acts analogously 
but yields a very different taste.}.

\begin{figure}[!ht]
\begin{center} 
\epsfig{file=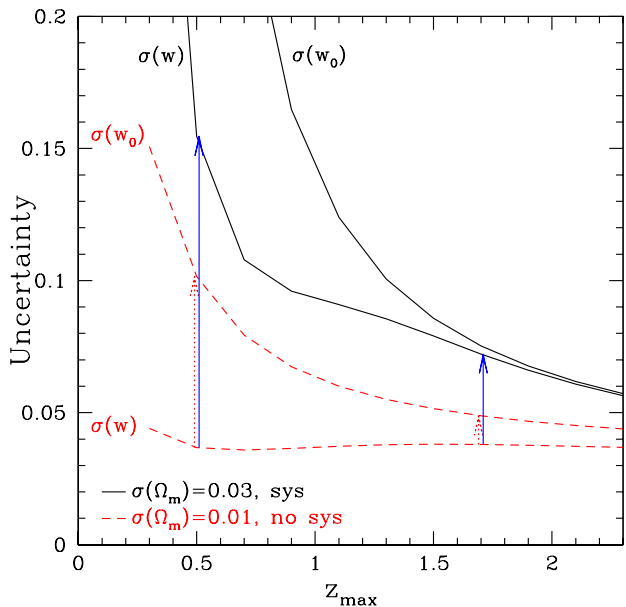, height=3.3in, width=3.4in} 
\caption{
Uncertainty in determination of the dark energy equation of state
today as a function of survey depth $z_{max}$; $w$ denotes assuming a
priori that there is no time variation while $w_0$ allows the
possibility.  The red, dotted arrows denote the difference; ignoring
the possibility that $w$ varies with time grossly underestimates the
error, especially for shallow surveys.  The blue, solid arrows show
the effect of ignoring systematic errors.  Precisely (and accurately)
determining the equation of state requires supernovae at $z>1.5$. }
\label{sigww0}
\end{center} 
\end{figure}

The mere possibility of time variation also carries important
implications for error estimation.  An a priori assumption of constant
behavior not only biases the conclusions on cosmology and dark energy,
but gives strongly deviant estimations of the associated errors,
illustrated in Fig.~\ref{sigww0}.  That is, one gets inaccurate
results extremely precisely! The error $\sigma(w)$ -- assuming a
constant equation of state -- disagrees with $\sigma(w_0$) -- merely
{\it allowing} for the possibility of time variation -- by a
factor 3 for a survey observing 2000 (plus 300 low $z$) SNe out to
$z_{max}=0.5$.  Another virtue of a deep survey to $z>1.5$ is that
this disagreement is only 25\% at $z_{max}=1.7$. This is shown by the
red dotted arrows.

\begin{figure}[!ht]
\begin{center} 
\epsfig{file=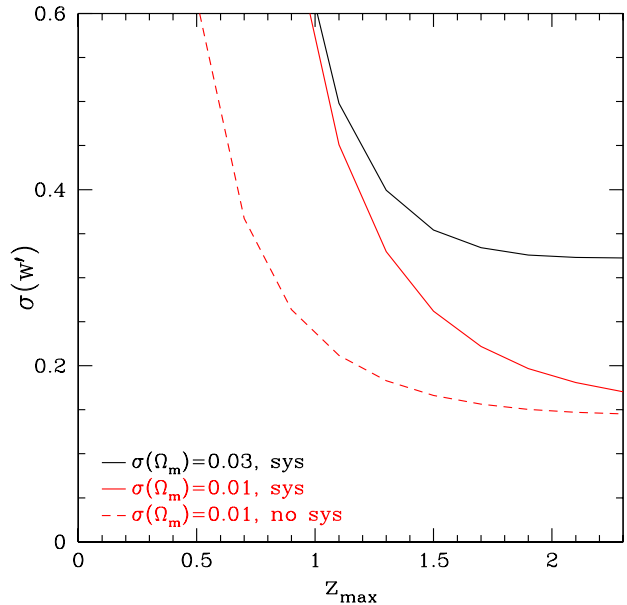, height=3.3in, width=3.4in} 
\caption{Uncertainty in determination of the time variation of the 
dark energy equation of state as a function of survey depth
$z_{max}$. Even in the idealized case of no systematic error the
uncertainty rises steeply as $z_{max}$ decreases.  One needs a survey
extending to $z_{max}\gtrsim 1.5$ to detect this key discriminator of
fundamental theories.  }
\label{sigw1}
\end{center} 
\end{figure}

The necessity for a long baseline survey is even more evident in 
Fig.~\ref{sigw1}, which  shows the uncertainty $\sigma(w')$.  The error 
sensitivity curve steepens dramatically as the depth decreases below 
$z_{max}=1.5$, rapidly worsening to uselessness. 

Along with the uncertainty in dark energy properties is that in our
cosmological knowledge.  So rather than fixing the dimensionless matter
density $\Omega_M$, we take as a realistic case a gaussian 
prior $\sigma(\Omega_M)=0.03$, i.e.~$\Omega_M=0.3\pm0.03$.

\section{Heresy by Deed: Ignoring Systematic Errors} 

Uncertainties in source, propagation, or detector impose a floor on
our ability to reduce errors merely by gathering large numbers of
supernovae.  While the great advantages of supernovae as a probe are
the long history of supernova studies, the rich data stream and
crosschecks they provide in their lightcurves and spectra, and their
underlying physical simplicity, we still cannot ignore the impact of
astrophysics on our attempts to measure cosmology.

In Fig.~\ref{sigww0} we see the huge discrepancy between the precision
claimed in the ideal situation (actually with a prior
$\sigma(\Omega_M)=0.01$, not fixed $\Omega_M$) and in the presence of
systematics (see blue solid arrows).  The systematic error essentially
represents imperfect knowledge of all the astrophysics lying behind
the observations, leaving a small residual error once we have carried
out as good a fit as possible to the data.  The systematic imposes an
upper limit on the number of supernovae useful for reducing the
statistical error in the magnitude through Poisson statistics.  One
example of such a systematic is nonstandard host galaxy dust
extinction.  To model the slow variation of astrophysical systematics
we adopted a floor to the magnitude error within a bin of width
$\Delta z=0.1$ of $dm=0.02\,(1.7/z_{max})\,(1+z)/2.7$.  Despite the
error growing with redshift, we see from Fig.~\ref{sigww0} that the
long baseline of a deep survey provides crucial leverage.

Indeed this conclusion might be made even stronger.  Despite an 
increased magnitude error for short redshift baselines, our adopted 
systematic might be said to be overly generous to shallow 
surveys (e.g.~it gives an error of 0.02 at $z=0.5$ for a survey 
reaching $z_{max}=0.9$), since the level of 
the residual systematic will depend on how elaborately the survey is 
designed.  Without a long redshift baseline, broad wavelength coverage 
into the near infrared, spectral observations, a rapid observing 
cadence, small point spread function, etc.~this number can be large. 
SNAP is specifically designed to achieve 0.02 mag.  For a typical ground 
based survey, a more realistic estimate might be 0.05 mag.  

For the time variation $w'$ in Fig.~\ref{sigw1} the discrepancy due to
ignoring systematics is also strong. For any reasonable prior on
$\Omega_M$, systematics have an extreme effect for shallow surveys: a
factor $\sim5$ degradation of our estimate $\sigma(w')$ at
$z_{max}=0.5$.  Compare this to a mere 12\% (40\%) degradation for
$z_{max}=1.7$ when the $\Omega_M$ prior is 0.03 (0.01); this clearly
shows the vast utility of including supernovae at $z>1.5$.

\section{Heresies Compound} 

We have seen that low redshift sensitivity to the form of the dark
energy depends on idealized conditions: 1) reduction of the parameter
space by fixing the cosmological model 
(i.e.~the matter density $\Omega_M$), 2) reduction of the parameter space by restricting
the dark energy model (i.e.~ad hoc adoption of constant $w$, ignoring
$w'$), 3) reducing errors by increasing statistics without limit 
(i.e.~no systematics floor from unknown uncertainties). This perfect
knowledge of cosmology, physics, and astrophysics is unrealistic and
misleading.

Compounding approximations takes us further from reality. 
Here we take the three oversimplifications two at 
a time to show the distortions they cause.  The conclusion in 
each case will be that realistic analysis of probing dark energy 
leads inexorably to the necessity for the observations to extend  
beyond $z>1.5$.

\begin{figure}[!ht]
\begin{center} 
\epsfig{file=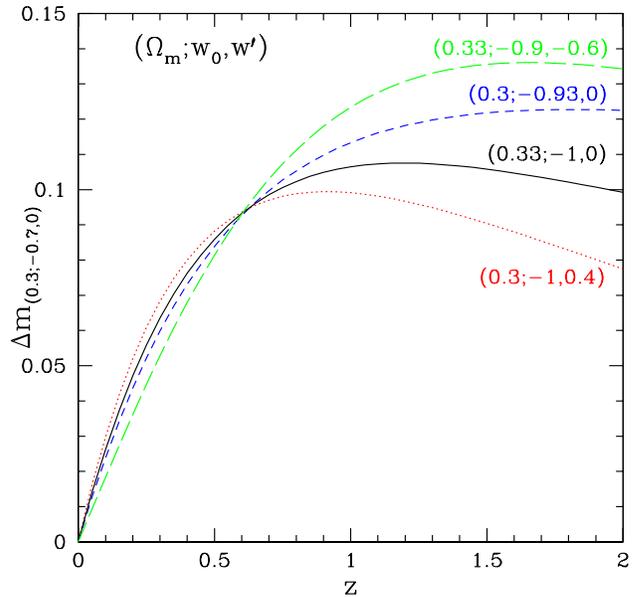, height=3.3in, width=3.4in} 
\caption{Degeneracies due to the dark energy model, e.g.~equation of 
state value $w_0$ or evolution $w'$, and to the cosmological model,
e.g.~value of $\Omega_m$, cannot be resolved at low redshifts.  In
this differential magnitude-redshift diagram the three parameters to
be determined are varied two at a time. Only at $z\approx1.7$ do these
very different physics models exceed 0.02 mag discrimination; SNAP
will be able to distinguish them. }
\label{degen}
\end{center} 
\end{figure}

For clarity and conciseness, we demonstrate this in simple 
illustrations.  Fig.~\ref{degen} shows the effects of correcting 
the first two oversimplifications.  When both $\Omega_M$ and the 
dark energy model (e.g.~constant $w$) are not overassumed, then degeneracies 
can lead to complete inability to discriminate very different cases 
using only data from a survey out to $z\le1$.  A deep survey gains 
both by the divergence of the curves and the longer redshift observation 
baseline.  The curves in Fig.~\ref{degen} would be distinguishable by 
SNAP, which will attain a precision, including  
systematics, below 0.02 mag. 

The effect of the second and third heresies is to mistake the uppermost, 
more realistic  
curve on Fig.~\ref{sigww0} for the lowest one. 
Ignoring both time variation and 
systematics would misestimate the errors by a 
factor 12.5 at $z_{max}=0.5$ but only 2 at $z_{max}=1.7$.  

\begin{figure}[!ht]
\begin{center} 
\epsfig{file=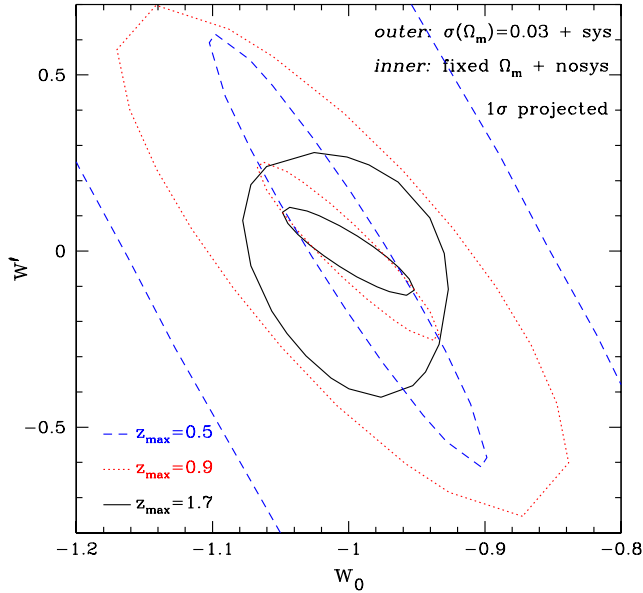, height=3.3in, width=3.4in} 
\caption{The effect of breaking oversimplifying assumptions on 
cosmological parameter determination as a function of survey depth 
$z_{max}$.  Uncertainties in $\Omega_m$ and the presence of 
systematics drastically weaken constraints from shallow surveys 
but 
the long baseline and depth $z_{max}>1.5$ 
immunize against 
systematics.  The outer contours of each of the three pairs 
represent realistic estimates for the cosmological parameters as a 
function of survey depth (see \cite{sys}). 
Contours here enclose 39\% of the probability so the 1$\sigma$ errors can 
be read off by projection onto the axes. 
} 
\label{ellsys}
\end{center} 
\end{figure}

Finally, consider the first and third together: the idealized case
vs.~realistic knowledge of the cosmology in the form of flatness, a
prior on $\Omega_M$ of 0.03, and systematic error. 
Fig.~\ref{ellsys} illustrates several important properties:
\begin{enumerate} 
\item $w'$: A shallow 
survey is incapable of appreciably limiting $w'$, even for  
perfect assumptions; a medium survey fails under any realistic conditions. 
\item Depth: While there appears to be only moderate  
difference between the results of a $z_{max}=0.9$ and 1.7
survey under the ideal case, for the realistic case the 1$\sigma$
constraints on $w_0$, $w'$ degrade by a full sigma. 
Depth plus long redshift baselines immunize against the effect of
systematics.  The main remaining influence is the degeneracy from an
uncertain $\Omega_M$, which can be dealt with by complementary
cosmological information (see the next section).\footnote{Note also 
that uncertainty in $\Omega_M$ tends to fatten contours in one 
direction. 
Especially for the shallow survey cases the limits on $w_0$, 
$w'$ change relatively little with increasing uncertainty on $\Omega_M$, 
but the area of the error contours increases by up to a factor three.  So 
one must be cautious at low redshift of simple quotes such as ``this 
experiment determines $w_0$ to $\pm0.1$''.} 
\item Like to like: Experiments should be compared under the 
appropriate assumptions.  An idealized $z=0.9$ survey might 
unfairly claim limits on 
$w_0$, $w'$ {\it better} than the realistic $z=1.7$ one, 
in noted contrast to the above like to like comparison. 
\end{enumerate}

\begin{figure}[!ht]
\begin{center} 
\epsfig{file=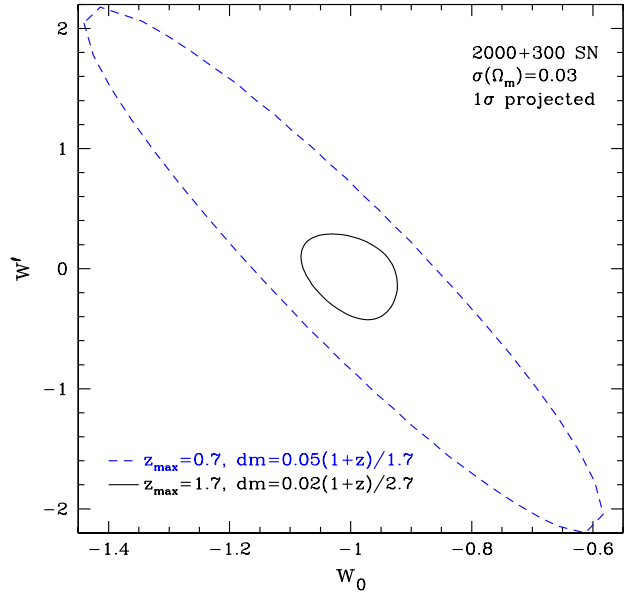, height=3.3in, width=3.4in} 
\caption{Realistic assessment of cosmological parameters from complete 
and precise surveys in the next decade from the ground 
($z_{max}=0.7$) and space ($z_{max}=1.7$) \cite{sys}.  
Contours here enclose 39\% of the probability so the 1$\sigma$ errors can 
be read off by projection onto the axes. 
} 
\label{ell717}
\end{center} 
\end{figure}

As a final wrap up, consider Fig.~\ref{ell717}.  This illustrates 
the comparison between surveys to $z_{max}=0.7$ and 1.7, roughly corresponding 
to the depths for completeness and precision from ground based and space 
based supernova surveys in the next decade.  Each includes 2000 supernovae 
plus an additional 300 at $z<0.1$, and makes realistic assumptions about 
cosmological and astrophysical knowledge.  The deep survey is seen to 
represent a huge advancement in determination of the dark energy model.

\section{Role of Complementary Probes} 

Complementary probes of cosmology such as the cosmic microwave
background (CMB), weak gravitational lensing, galaxy counts, etc.~play
an important role in elucidating dark energy. In particular, they are
crucial for constraining flatness and the matter density $\Omega_M$.
They will also impact, together with supernovae and perhaps
independently, the determination of a redshift averaged form of the
equation of state $\langle w\rangle$.  But these probes possess very
little sensitivity to the physically decisive time variation $w'$, and
even any prior constraint provided on $\langle w\rangle$ contributes
minimally to finding $w'$. Furthermore, except for the CMB (which does
not see time variation since it measures the distance to a single
redshift), they are first generation experiments, with their own
systematic effects (over the 2/3 of the age of the universe stretching
back to $z\approx1.5$) at best partially accounted for.

Several supernova cosmology surveys will go forward over the next
several years. For example, the ``$w$ Project'' \cite {wproj} at CTIO
should obtain 200 SN at redshifts $z=0.15-0.75$ over the course of
five years. With a quoted systematic \cite{riess} of $dm=0.03(z/0.5)$,
and using a prior of $\sigma(\Omega_M)=0.04$ and the crucial low 
redshift data of the Nearby Supernova Factory \cite{snf}, this should 
determine $w$ to $+0.10,-0.12$.  
Suppose $\sigma(\langle w\rangle)=0.1$, where $\langle w\rangle$ is
interpreted as an average value of the EOS over the redshift range.
This would of course be quite interesting in itself, but for the
further important parameter $w'$ such middle redshift experiments
provide no useful prior.  In fact, such a prior on $\langle w\rangle$
would improve SNAP's constraint on $w'$ by less than 3\%.  In
this sense SNAP is very much a next generation experiment.

One promising method of adding value to SNAP is the information 
the Planck Surveyor experiment~\cite{Planck} provides via the cosmic 
microwave 
background anisotropies.  This constrains a combination of the 
matter density and the dark energy parameters; the result of this 
complementarity is not only to strengthen the advantage of a high 
redshift supernova survey, but to greatly improve its precision 
\cite{fhlt}. 
For example, adding the information expected from Planck 
would improve SNAP's determination of $w'$ by
roughly a factor of two.  In fact, using a new, well behaved 
parametrization of the function $w(z)$,
Linder \cite{Li02} shows that one could attain 
$\sigma(dw/d\ln(1+z)|_{z=1})\approx0.1$ for a model such as supergravity 
inspired dark energy.  For the particular SUGRA model
\cite{braxm} this would represent a 99\% confidence level detection of time
variation in the EOS.

\section{Conclusion} 

The discussions and illustrations presented here 
show that expectations based on oversimplified 
cosmology, physics, and astrophysics prove insufficient and 
misleading for understanding how to probe dark energy.  
Could we detect dark energy with measurements at $z<1$?  Assuredly -- 
we already have through the supernova method.  Could we reliably 
distinguish its equation of state from that of a cosmological 
constant?  Possibly -- wide field ground based surveys, possibly 
together with higher redshift Hubble Space Telescope observations, 
could well give indications of this, though not necessarily definitive 
ones.  Could we see the critical evidence of time variation in the 
equation of state that sets us on the path of a fundamental 
theory?  No.  For that we required detailed observations out to 
$z\approx1.5-2$ and control of systematics. 

In the realistic view, one clearly appreciates the need 
for a precision survey reaching out to $z_{max}\approx 1.5-2$. 
More rigorous Monte Carlo simulations \cite{sys} implementing a variety 
of systematic error, cosmology, and dark energy models bear 
out this conclusion.  

\vspace{0.2in}

\section*{Acknowledgments}

This work was supported at LBL by the Director, Office of Science, 
DOE under DE-AC03-76SF00098 and at CWRU by a DOE grant 
to the particle astrophysics theory 
group.  We gratefully acknowledge Ramon Miquel and Nick
Mostek for computations.

\end{document}